\documentclass[oldversion]{aa} 
\usepackage{graphicx}
\usepackage{txfonts}
\usepackage{natbib}
\bibpunct{(}{)}{;}{a}{}{,} 

\begin{document}
   \title{Old stellar counter-rotating components in early-type galaxies \\from elliptical-spiral mergers }
   \author{P. Di Matteo
          \inst{1}
          \and
          F. Combes\inst{1}
          \and
          A.-L. Melchior\inst{1,2}
          \and
          B. Semelin\inst{1,2}
          }


   \offprints{P. Di Matteo, \email{paola.dimatteo@obspm.fr}}

   \institute{LERMA, UMR 8112, CNRS, Observatoire de Paris, 61 Avenue de l'Observatoire, 75014 Paris, France \and
Universit\'e Pierre et Marie Curie - Paris 6, 4 Place Jussieu, 75252 Paris Cedex 5, France
             }

   \date{... ; ...}

 
  \abstract
   {We investigate, by means of numerical simulations, the possibility of forming counter-rotating old stellar components by major mergers between an elliptical and a spiral galaxy. We show that counter-rotation can appear both in dissipative and dissipationless retrograde mergers, and it is mostly associated to the presence of  a disk component, which preserves part of its initial spin. In turn, the external regions of the two interacting galaxies acquire part of the orbital angular momentum, due to the action of tidal forces exerted on each galaxy by the companion. }


   \maketitle
%

\section{Introduction}

Simulations show that stellar counter-rotation in galaxies could emerge thanks to two different processes: dissipative and dissipationless mergers.\\
Confirming the suggestion  of \citet{korm84}, \citet{balquin90}
showed that unequal mass mergers of elliptical galaxies 
 can produce counter-rotation if the orbit of the encounter is retrograde with respect to the spin of the primary. They also pointed out that the rotation seen in the counter-rotating component is a tracer of the orbital angular momentum and that  both primary and secondary stars counter-rotate at the core.\\
\citet{hb91} showed that counter-rotating central gas disks can form as a result of retrograde mergers between  two gas rich spiral galaxies, discussing the possibility that star formation in such disks could produce components with decoupled kinematics, as in the core of some elliptical galaxies.\\
Some years later, \citet{balgon98} showed that kinematically peculiar cores may be generated also in retrograde stellar spiral-spiral mergers: in this picture, the central bulges transport orbital angular momentum inward to the center of the remnant, while the outer parts keep the spin signature of the precursor disks. 
Also \citet{bendob00} put in evidence the possibility to form counter-rotation at large radii, simulating mergers between equal mass disk galaxies.
Finally, \citet{jes07}, studying the 2D kinematics of a sample of simulated disk merger remnants, showed that counter rotating cores, made of old stellar populations, are almost exclusively formed in equal mass mergers where a dissipative component is included. 
Evidently, both  mechanisms (the dissipative and dissipationless one) can occur in real systems, producing a variety of kinematically decoupled components, of different ages and physical extensions \citep{mc06}.

In this paper, we want to present a new scenario, according to which stellar counter-rotation can form both in dissipative and dissipationless retrograde\footnote{i.e. the spin of the spiral galaxy is initially anti-parallel to the orbital angular momentum.} major mergers of elliptical-spirals galaxies\footnote{Note that all the previous  numerical works refer to mergers between two ellipticals or two disk galaxies. So far, the possibility of forming counter-rotating cores by mergers of galaxies of different morphologies has never been exploited.}. In this scenario, stars in the external regions of the galaxies involved in the encounter acquire orbital angular momentum, at first pericenter passage and, mostly, in the last phases of the merging process, while those in the most inner regions can maintain some of the initial spin (anti-parallel to the orbital one), producing decoupled counter-rotating central components. In the case of dissipative mergers, the central decoupled core could be composed of two different populations: the old stellar population, which has preserved its initial spin, and a new stellar population, born \emph{in situ} from the kinematically decoupled gas component. The redistribution of the angular momentum between the different galactic components (gas, stars and dark matter) is  discussed.

\section{The simulations}\label{simul}
The simulations examined here are a small subset of the simulations described in \citet{dimatteo07}. In that paper, we presented a library of hundreds of galaxy encounters, involving systems with 1:1 mass ratios, and of different morphologies. 
For a description of  the numerical code adopted to run the simulations, we refer the reader to the above  paper. \\
In particular, in the following, we will study the velocity field of a galaxy, remnant of a merger between an elliptical and a spiral Sa (hereafter called gE0 and  gSa, respectively). These results will be compared to those obtained for a merger between an elliptical and a disk galaxy without gas (hereafter called gS0),  when identical orbital initial conditions are chosen. The comparison between these two experiments will allow to determine the influence of the dissipative component (gas) in the formation of the counter-rotating core.  The main parameters of the simulated galaxies are summarized in Table 1.  The stellar component of the elliptical galaxy has been modelled as a Plummer sphere. As we checked, the surface density profile of the adopted distribution well reproduces a de Vaucouleurs $R^{1/4}$ law, for $4 \rm{kpc} \le r \le 13\rm{kpc}$.\\ In the two cases studied here
, the initial separation between the two galaxies is 100 kpc. The spin of the disk galaxy is antiparallel to the orbital angular momentum 
, while the elliptical galaxy does not initially rotate. The relative distance between the two systems at first pericenter passage is 8 kpc and their relative velocity\footnote{The values refer to the ideal Keplerian orbit of two equal point masses of mass $m=2.3\times 10^{11} \rm{M_\odot}$} is 707 km/s.  These two cases have been chosen because they well represent the formation mechanism of counter-rotating cores that we want to describe here. In fact, in all the simulations involving an elliptical and a spiral Sa galaxy on retrograde orbits analysed so far, we found the presence of a counter-rotating core in the remnant galaxy. These cases will be discussed in Sect. \ref{app1}. They involve  interactions where the relative inclination between the Sa disk and the orbital plane is $0\le i\le 20^0$.

\section{Results}
This section describes in detail the formation of counter-rotating systems. In particular, Sect.\ref{e0sa} will deal with the analysis of the kinematics of the remnant galaxy, seen edge-on, 1 Gyr after the retrograde dissipative merger. 
We will discuss the 2D velocity maps of the remnant (Sect.\ref{vlosmaps}), showing the presence of an old stellar counter-rotating core (Sect.\ref{old}). To understand the formation mechanism of this decoupled component, in Sect.\ref{moment}, we will analyze and discuss the evolution of the total and specific angular momenta, during the encounter. Finally, in Sect.\ref{e0s0}, these results will be compared to those obtained in the case of a dissipationless merger. 

\subsection{A dissipative retrograde encounter between an elliptical and a spiral Sa}\label{e0sa}

\subsubsection{Line-of-sight velocity maps}\label{vlosmaps}

The 2D velocity maps of the remnant galaxy are shown in Fig.\ref{vrtot}, for the gas\footnote{In fact, the gas component contains also the new stars, formed during the interaction.}, old stars and dark matter component. When looking at these maps, some peculiar features clearly appear:  gas and new stars counter-rotate with respect to the main stellar body of the galaxy (respectively, left and central panels in Fig.\ref{vrtot}); the stellar component shows the presence of a central decoupled counter-rotating region, which is made by \emph{old stars}; the amount of rotation found in the dark matter component (right panel in  Fig.\ref{vrtot}) is comparable to that found for the stellar main body.

\subsubsection{Counter-rotation in the old stellar component} \label{old}
\begin{figure}
   \centering
  \includegraphics[width=8cm,angle=0]{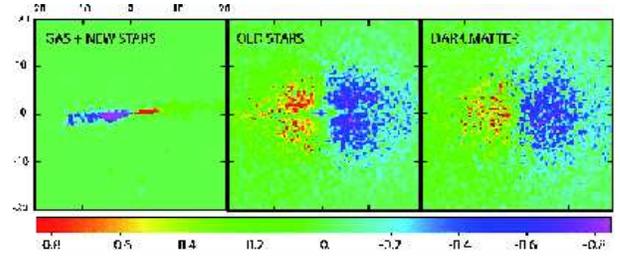}
   \caption{Line-of-sight velocity maps of gas (left panel), old stars (central panel) and dark matter (right panel) of the remnant of an elliptical-spiral merger. The maps are evaluated  1Gyr after the coalescence of the two galaxies.  Each side of the plot is 40 kpc in size. Velocities are in units of 100 km/s.}
              \label{vrtot}%
    \end{figure}

\begin{figure}
   \centering
   \includegraphics[width=6.cm,angle=0]{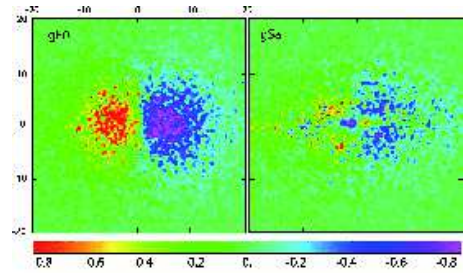}
   \caption{Line-of-sight velocity maps for the old stellar component of the remnant  of an elliptical-spiral Sa merger. Left panel: stars from gE0 initially; right panel: stars initially in gSa. Each side of the plot is 40 kpc in size. Velocities  are in units of 100 km/s. The counter-rotating region is clearly visible in the central part of the map on the right.}
              \label{vrstar}%
    \end{figure}

\begin{figure}
   \centering
   \includegraphics[width=3.5cm,angle=270]{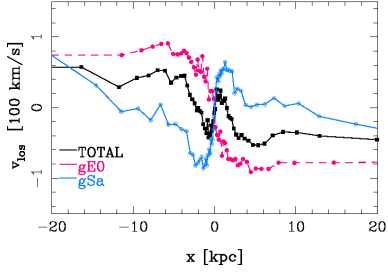}
   \caption{Rotation curve  of the old stellar component of an elliptical-spiral Sa merger. Black curve: line-of-sight velocity profile of the total old stellar component; red curve: line-of-sight velocity profile of stars initially belonging to the elliptical galaxy; blue curve: line-of-sight velocity profile of stars initially belonging to the spiral Sa.}
              \label{vlosgas}
    \end{figure}

But where does the counter-rotation in the old stellar component come from?
To answer this question, we separate the stellar velocity map shown in Fig. \ref{vrtot} into two contributions, that due to the stars initially (i.e. at time t=0 of the simulation) belonging to the elliptical and the one due to the stars initially populating the spiral Sa. The resulting 2D velocity maps are presented in Fig. \ref{vrstar}. They clearly show that the counter-rotating region is completely associated to stars initially belonging to the spiral. In other words, the old stellar component, initially belonging to the spiral Sa, contains a decoupled counter-rotating central region, which is not present among stars initially belonging to the elliptical.
This feature is still more evident when looking at the rotation curve of the remnant (Fig. \ref{vlosgas}). 

   \begin{table}
      \caption[]{Galaxy parameters. The bulge and the halo are modelled initially as Plummer spheres, with characteristic masses $M_B$ and $M_H$ and characteristic radii $r_B$ and $r_H$.  $M_{*}$ and  $M_{g}$ represent the masses of the stellar and gaseous disks, whose vertical and radial scale lengths are given, respectively, by $h_{*}$, $a_{*}$, and $h_{g}$, $a_{g}$. Masses are in units of $2.3\times10^9 \mathrm{M_{\odot}}$ and distances are in kpc.}      
         \label{galpar}
	 \centering
         \begin{tabular}{ccccccccccc}
            \hline\hline
	    &  $M_{B}$&$M_{H}$&$M_{*}$&$M_{g}/M_{*}$&$r_{B}$&$r_{H}$&$a_{*}$&$h_{*}$&$a_{g}$&$h_{g}$\\
            \hline
	    gE0 &  70&30&0&0&4&7&--&--&--&-- \\
	    gSa & 10&50&40&0.1&2&10&4&0.5&5&0.2 \\
	    gS0 & 10&50&40&0.&2&10&4&0.5&--&-- \\
            \hline
         \end{tabular}
   \end{table}


To evaluate this profile, we chose a slit parallel to the galaxy major axis, being the galaxy seen edge on, as in Fig.\ref{vrtot}. The width of the simulated slit is 2 kpc, and it is symmetrically placed with respect to the galaxy vertical direction.  The rotation curves in this figure show 
 that the counter-rotation found in  Fig.\ref{vrtot} is completely due to stars belonging, at t=0, to the spiral galaxy.
   \begin{figure}
   \centering
   \includegraphics[width=4cm,angle=270]{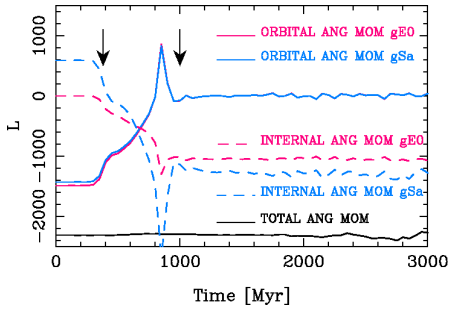}
   \caption{Evolution, with time, of the angular momentum $L$ during an elliptical-spiral Sa merger. Black line: total angular momentum; solid colored lines: orbital angular momentum of the elliptical (red curve) and of the Sa galaxy (blue curve); dashed colored lines: internal angular momentum of the elliptical (red curve) and of the Sa galaxy (blue curve). The angular momentum is in units of $2.3\times 10^{11} M_{\odot}\rm{ kpc kms^{-1}}$. The two arrows indicate the first pericenter passage and the merging time.}
              \label{angmomgas}%
    \end{figure}

\begin{figure}
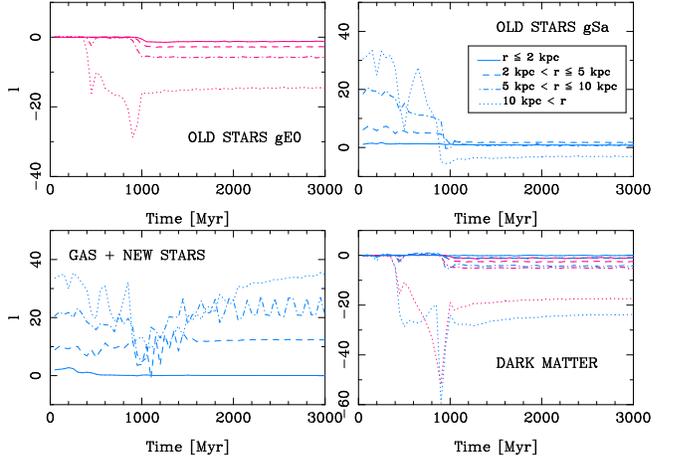

 \begin{minipage}[b]{3.5cm}
   \centering
   \includegraphics[width=3cm,angle=270]{dimatteo5.ps}
 \end{minipage}
 \ \hspace{2mm} \hspace{3mm} \
 \begin{minipage}[b]{3.5cm}
  \centering
   \includegraphics[width=3cm,angle=270]{dimatteo6.ps}
 \end{minipage}\\
 \ \hspace{2mm} \hspace{3mm} \
 \begin{minipage}[b]{3.5cm}
  \centering
   \includegraphics[width=3cm,angle=270]{dimatteo7.ps}
 \end{minipage}
 \ \hspace{2mm} \hspace{3mm} \
 \begin{minipage}[b]{3.5cm}
  \centering
   \includegraphics[width=3cm,angle=270]{dimatteo8.ps}
 \end{minipage}
   \caption{Evolution, with time, of the specific angular momentum $l$ for different regions and different components of the elliptical and Sa galaxies. In each panel, four regions are shown, as explained in the legend. 
Upper left panel: specific angular momenta of the old stellar component belonging initially to the elliptical galaxy; upper right panel: specific angular momenta of the old stellar component belonging initially to the Sa galaxy; lower left panel: specific angular momenta of the gas component of the spiral galaxy; specific angular momenta of the dark matter component of the elliptical  galaxy (red curves) and of the spiral Sa (blue curves). The specific angular momentum is in units of $10^2 \rm{ kpc kms^{-1}}$.}\label{zone}
\end{figure}
   \begin{figure}
   \centering
   \includegraphics[width=7cm,angle=0]{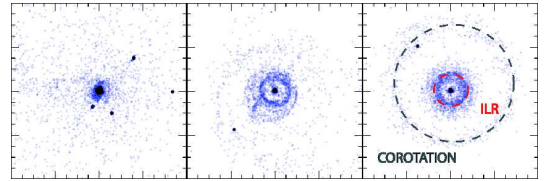}
   \caption{From left to right: gas maps at  t=1 Gyr (corresponding to the merging time), t=2 Gyr, and t=3 Gyr. Each side of the plot is 40 kpc in size. The presence of two rings is clearly visible in the last configuration, at t=3 Gyr.}
              \label{gasmap}%
    \end{figure}

 It also shows that the counter-rotation in the stellar component associated to the Sa galaxy is very extended (about 10 kpc in radius). In turn, the stellar component associated to the elliptical does not show counter-rotation, but it rotates in the opposite direction with respect to the decoupled core (i.e., parallel to the remnant main body). Finally, the relative distribution of stars initially belonging to the elliptical and to the spiral galaxy determines the total rotation curve of the old stellar component shown in Fig.\ref{vlosgas}. Indeed, as we checked, stars initially located in the elliptical galaxy dominate the density distribution at large radii, while, in turn, those initially belonging to the spiral dominate in the central regions (see Appendix \ref{app2} for a discussion on this point). The total counter-rotating region is obviously due to the superposition of these two contributions,  resulting into a less extended (about 2 kpc in radius) region.\\

\subsubsection{Evolution of the total and specific angular momentum}\label{moment}

The aim of this section is to understand which is the mechanism that can produce the counter-rotation in the old stellar component found in Figs.\ref{vrtot} and \ref{vlosgas}. In other words, how the angular momentum is redistributed between the different galactic components during the interaction?\\
To answer this question, we, first of all, analyzed  the evolution of the total angular momentum $L_{tot}$, as shown in Fig.\ref{angmomgas} (black line). During the simulation, the total angular momentum is conserved with an accuracy of 2\%. The total number of steps of the simulation being $N_{steps}=6000$, this  means that the relative error in the total angular momentum is of the order of $10^{-6}$ per step. Initially this total angular momentum is distributed mainly in the orbits of the two galaxies, and as internal spin of the spiral Sa.
This situation keeps unchanged up to the first pericenter passage between the two systems: at this time (t=380 Myr), the gravitational torques exerted by each galaxy on the companion redistribute part of the initial orbital angular momentum into internal spin of the two systems. The result is that both galaxies acquire part of the (negative) orbital spin: the elliptical begins to rotate with a spin parallel to the initial orbital angular momentum and the spiral loses part of its initial spin, which was anti-parallel to the orbital one. The transfer of angular momentum from the orbit to the internal spin of the two galaxies, via tidal torques, continues until the two systems approach the final stages of the merger event (t=1 Gyr). At this time, the rapidly changing tidal forces exert positive and negative torques on the companion (see the peaks at t=820 Myr in the evolution of the orbital and internal angular momenta in Fig.\ref{angmomgas}). Soon after, the two galaxies merge, and the total angular momentum is completely distributed into the galaxies, as internal rotation (dashed lines in Fig.\ref{angmomgas}).  \\
The next step in this picture is to understand how the angular momentum is redistributed among the gas, stars and dark matter during the interaction. This study is summarized in Fig.\ref{zone}, where the temporal evolution of specific angular momentum  is shown, for each of these components. To better understand the emergence of a counter-rotating region in the old stellar component associated to the spiral galaxy, we evaluated the specific angular momentum of gas, stars and dark matter particles, separating the contribution from the elliptical and the spiral galaxies, and analysing it for four different regions of each of the two galaxies ($r\le 2 \rm{kpc}$, $2 < r \le 5\rm{kpc}$, $5 < r \le 10\rm{kpc}$, and $r > 10 \rm{kpc}$, with $r$ the distance from the galaxy centers). The results in Fig.\ref{zone} show that all the galactic components acquire some of the initial orbital angular momentum. This is particularly true for stars and dark matter, where the acquisition of the orbital spin proceeds from the most external regions to the inner ones. Indeed, since the tidal forces are more important in the outer parts, the redistribution of the angular momentum affects more these outer regions.
 In the elliptical galaxy, for example, after the first pericenter passage (t=380 Myr), only stars at a distance $r> 10 \rm{kpc}$ have acquired some internal spin, while the inner ones show no signs of rotation. As the interaction goes on, towards the final merging phase, tidal torques begin to affect inner regions. For the spiral, while these torques are sufficient to reverse the initial spin at distances greater than $10 kpc$, in turn, they are not strong enough to reverse this spin in the internal regions, which ultimately continue to rotate with a spin parallel to the initial one. This gives rise to the emergence of a counter-rotating region, among stars initially belonging to the spiral.  For the gas, the specific angular momentum  in the external regions shows a minimum around t = 1Gyr (merging time), and increases later on. This feature is related to the formation of a bar in  the remnant disk. As we checked, this bar forces the incoming gas to redistribute along two  rings, coinciding with the inner Lindblad resonance and with the corotation radius. The formation of these structures are clearly visible in the gas maps in Fig.\ref{gasmap}.

\subsubsection{Counter-rotating region: bulge or disk stars?}\label{bulge}
Studying the formation of kinematic peculiar cores in stellar mergers of spiral galaxies, \citet{balgon98} showed
that the kinematical segregation appears because the central bulges transport orbital angular momentum inward, while the outer parts rotate according to the initial orientation of the precursors'spin. 
Separating the contribution from disk and bulge stars belonging initially to the spiral galaxy
, we find an alignment of the final rotation of the bulge material with the initial orbital angular momentum, as in \citet{balgon98}. But  \emph{contrary to these authors, we find that  the external regions also rotate according to the initial orientation of the orbital angular momentum}, as they are  more prone to acquire part of the orbital angular momentum via tidal torques. Consequently, the counter-rotating region (which has a spin anti-parallel to the initial orbital momentum) is mostly made of old stars initially belonging to the disk of the spiral rather than to its bulge. This is confirmed in Fig.\ref{angmombulge}, where the line-of-sight velocities of bulge and disk stars initially belonging to the spiral Sa galaxy are shown. 

\subsection{A dissipationless retrograde encounter between an elliptical and a disk galaxy}\label{e0s0}

   \begin{figure}
 \begin{minipage}[t]{4.1cm}
   \centering
    \includegraphics[width=3.cm,angle=270]{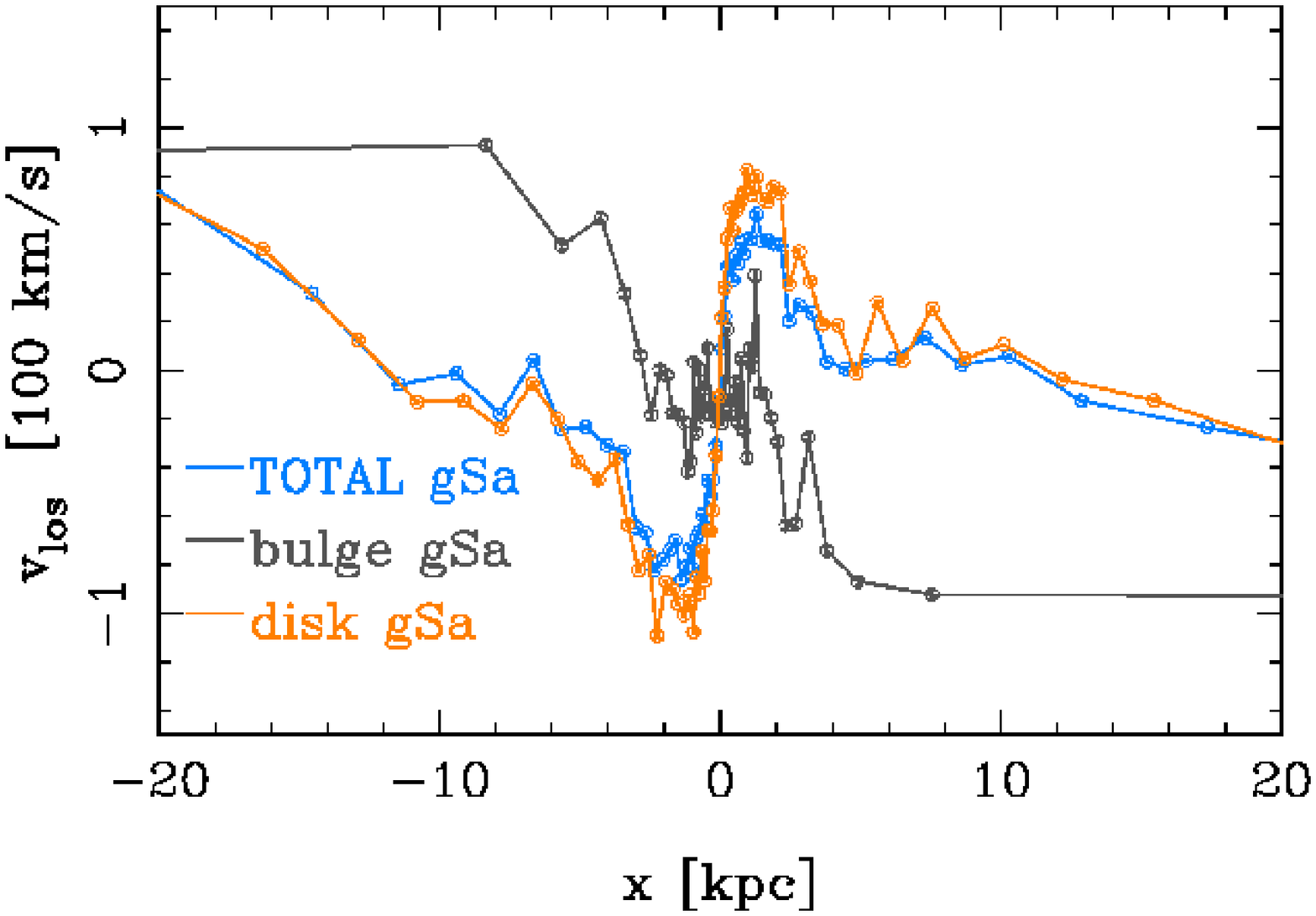}
   \caption{Rotation curve of the old stellar component, initially belonging to the spiral Sa galaxy, 1 Gyr after the coalescence of this system with an elliptical galaxy. Blue curve: total rotation curve; grey curve: rotation curve of stars initially belonging to the bulge; orange curve: rotation curve of stars initially belonging to the disk of the spiral. }
              \label{angmombulge}%
 \end{minipage}
 \ \hspace{1mm} \
\begin{minipage}[t]{4.5cm}
 \includegraphics[width=3.cm,angle=270]{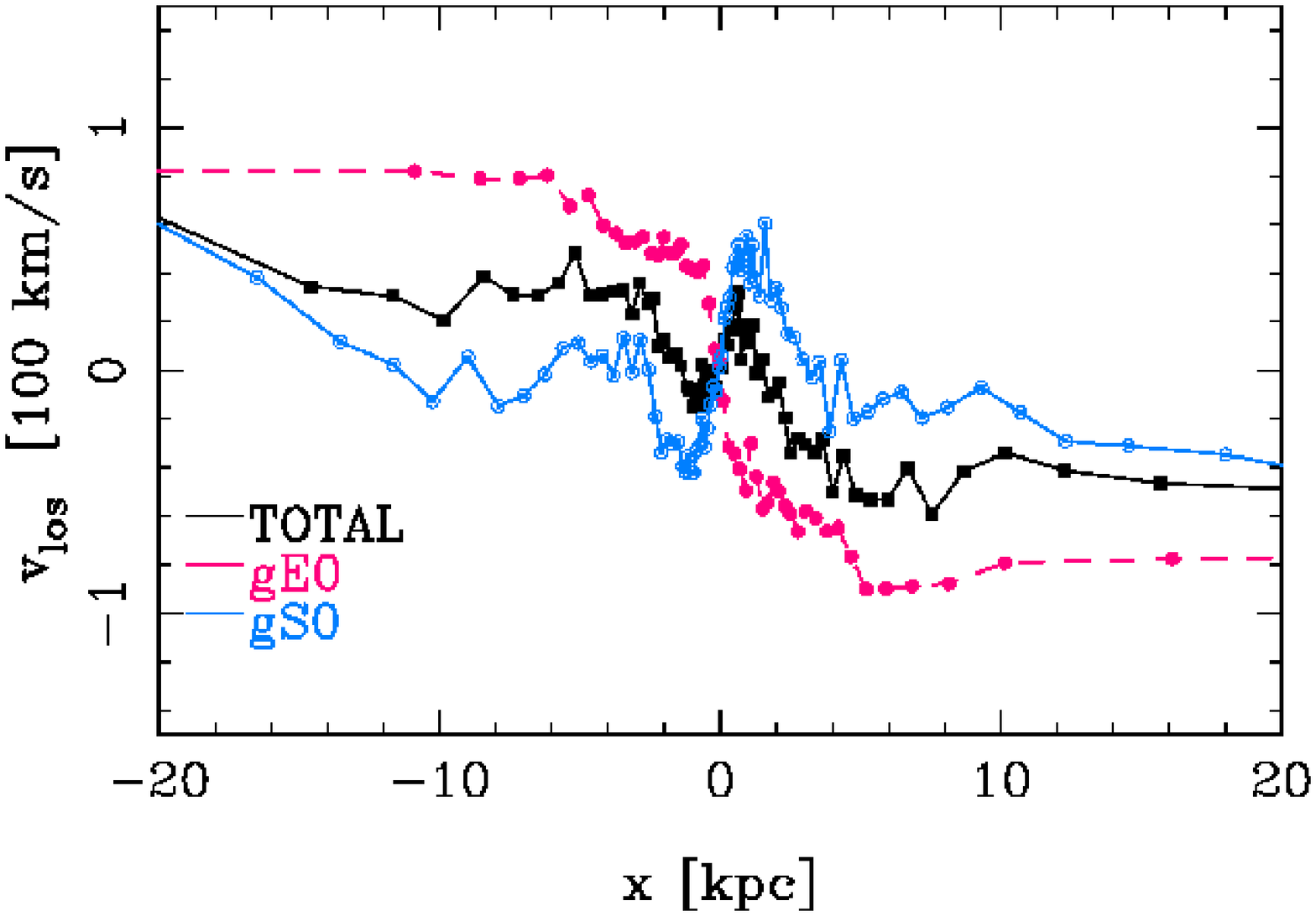}
   \caption{Rotation curve of the old stellar component of an E0-S0 merger, 1 Gyr after the coalescence of the two systems. Black curve: line-of-sight velocity profile of the total old stellar component; red curve: line-of-sight velocity profile of stars initially belonging to the elliptical galaxy; blue curve: line-of-sight velocity profile of stars initially belonging to the spiral S0.}
              \label{vrstarnogas}
 \end{minipage}
    \end{figure}
    
In the previous sections, we have shown the existence of a counter-rotating region completely associated with the old stellar component of the remnant galaxy. It is natural to ask whether and to what extent the presence of a dissipative component affects the emergence of a decoupled region and its extension. To this aim, we repeated the same simulation, adopting the same initial orbital conditions, but this time we did not include the gas component in the disk galaxy (see Table \ref{galpar}). The results are shown in Fig.\ref{vrstarnogas}. 
 They clearly show that the presence of a dissipative component is not a necessary condition to form a counter-rotating central core. Indeed, the rotation curve in Fig.\ref{vrstarnogas} shows that a decoupled central region is still present even if the disk galaxy does not contain gas. In turn, the presence of gas can modify the extension of this decoupled component:  when gas is not present, a larger amount of orbital angular momentum is acquired by the old stellar population in the inner regions. This results into a smaller extension of the counter-rotating component associated with the disk galaxy, and so, ultimately, with a smaller extension of the resulting total counter-rotating component of the merger remnant.



\subsection{Investigating other orbital parameters}\label{app1}

\begin{figure}
   \centering
   \includegraphics[width=7cm,angle=0]{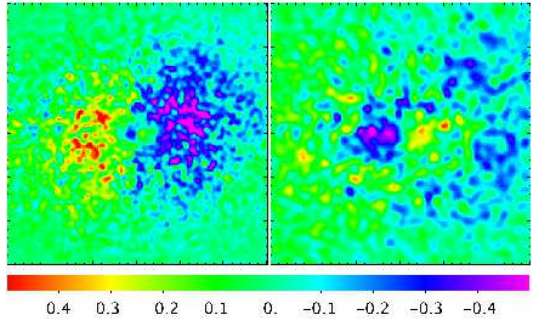}
   \caption{Line-of-sight velocity maps of the old stellar component of the remnant of a coplanar retrograde merger between an elliptical and a spiral Sa. The relative distance, at first pericenter passage, between the two galaxies is 8 kpc and their relative velocity is 742 km/s.  The maps are evaluated about 600 Myr after the coalescence of the two systems. In the left panel, each side of the plot is 40 kpc in size. The right panel shows a zoom of the inner central region (each side of the plot being 10 kpc in size). Velocities are in units of 100 km/s.} 
              \label{fig1app}%
\end{figure}

\begin{figure}
   \centering
   \includegraphics[width=7cm,angle=0]{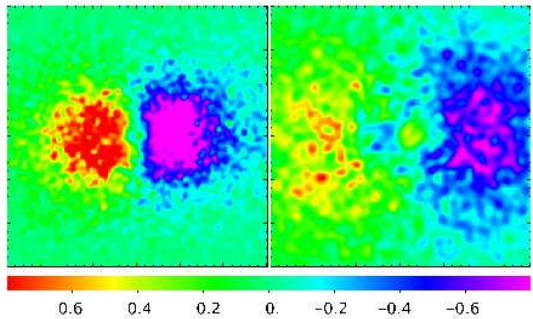}
   \caption{Same as Fig.\ref{fig1app}, but this time the maps refer to a coplanar encounter between two systems having a relative distance at first pericenter passage of 16 kpc and a relative velocity of 500 km/s. The maps are evaluated about 400 Myr after the coalescence of the two galaxies.}
              \label{fig2app}%
\end{figure}

\begin{figure}
   \centering
   \includegraphics[width=7cm,angle=0]{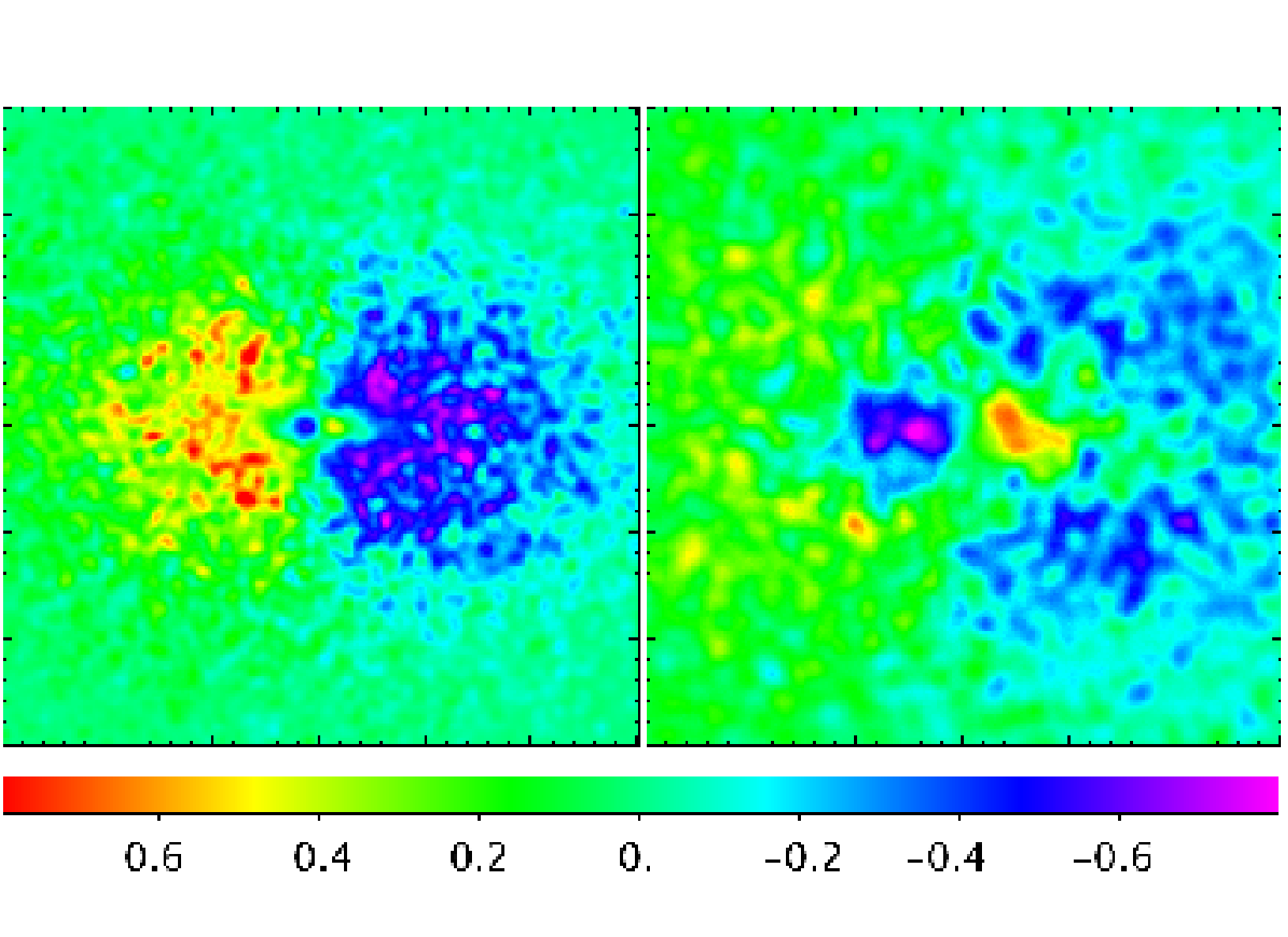}
   \caption{Same as Fig.\ref{fig1app}, but this time the maps refer to an encounter with $i=10^0$, being $i$ the inclination between  the spiral disk and the orbital plane.  The  two galaxies have a relative distance at first pericenter passage of 8 kpc and  a relative velocity of 707 km/s. The maps are evaluated about 1 Gyr after the coalescence of the two galaxies.}
              \label{fig3app}%
\end{figure}

\begin{figure}
   \centering
   \includegraphics[width=7cm,angle=0]{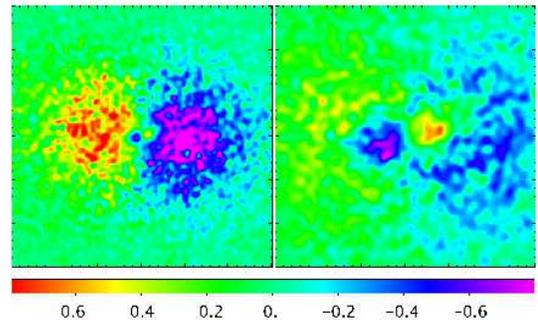}
   \caption{Same as Fig.\ref{fig1app}, but this time the maps refer to an encounter with  $i=20^0$, being $i$ the inclination between  the spiral disk and the orbital plane.  The  two galaxies have a relative distance at first pericenter passage of 8 kpc and  a relative velocity of 707 km/s. The maps are evaluated about 1 Gyr after the coalescence of the two galaxies.} 
              \label{fig4app}%
\end{figure}

As stated in Sect.\ref{simul},  counter-rotating central components were found in all the remnants of retrograde mergers involving an elliptical and a spiral Sa analysed so far.\\
In Figs. \ref{fig1app}, \ref{fig2app}, \ref{fig3app} and \ref{fig4app}, we present  2D velocity maps of some of these  remnant galaxies. Fig.\ref{fig1app} shows the velocity field of the remnant of a coplanar encounter between a E0 and a Sa galaxy,  having, at the first pericenter passage, a relative distance of 8 kpc (the same as the fiducial case described in Sect.\ref{e0sa}), and a relative velocity of 742 km/s.  The figure clearly shows that a decoupled counter-rotating region, whose extension is similar to the one of the fiducial dissipative simulation (Sect.\ref{e0sa}), is still present. A counter-rotating core is present also in Fig.\ref{fig2app}. In this case the remnant originates from a coplanar retrograde encounter between two systems, having a relative distance at pericenter of 16 kpc and a relative velocity of 500 km/s. In this case, the counter-rotating component is less extended (about 1 kpc), but the mechanism of its formation is the same described in Sect.\ref{moment}.\\
Counter-rotating components can also form in remnants of elliptical-spiral Sa encounters, when the disk of the spiral galaxy has a not null relative inclination respect to the orbital plane, as in the case shown in Fig.\ref{fig3app}. Note that, as the relative inclination $i$ increases, the velocity field of the remnant galaxy still presents  a counter-rotating component, but misaligned with the rotation axis of the main body (an example is given in Fig.\ref{fig4app}, which refers to an encounter with $i=20^0$).\\

\section{Conclusions}

We have presented a new scenario to form counter-rotating central components in early-type galaxies, by dissipative and dissipationless mergers of elliptical-spiral systems in retrograde orbits. 
In the case of dissipative mergers, the central decoupled core could be composed of two distinct populations: the old stellar population, which has preserved part of its initial spin, and a new stellar population, born \emph{in situ} from the kinematically decoupled gas component. Counter-rotating cores are also found in the remnants of non-coplanar mergers, at least for inclinations $i \le 20^0$ (being $i$ the angle between the spiral disk and the orbital plane).
We plan to realise a wider and  systematic study of the role played by orbital parameters and morphology of the interacting systems in the near future, varying the morphological parameters of the interacting galaxies (bulge-to-disk ratio of the spiral galaxy as well as different initial models for the elliptical), and the orbital initial conditions. 

\begin{acknowledgements}
The authors wish to thank the anonymous referee for her/his comments and instructive suggestions.\\ This research used computational resources of the Informatic Division of the Paris Observatory, and those available within the framework of the Horizon Project (see \emph{http://www.projet-horizon.fr/}).
2D velocity maps were produced using the python parallelized pNbody package, made by Y. Revaz (see \emph{http://aramis.obspm.fr/$\sim$revaz/pNbody/index.html}), and the SAOImage DS9  astronomical imaging and data visualization application.
\end{acknowledgements}


\begin{appendix} 
\section{Stellar profiles of the merger remnants: contribution from the former elliptical and spiral galaxies }\label{app2}

\begin{figure}
 \begin{minipage}[t]{9cm}
   \centering
   \includegraphics[width=5cm,angle=270]{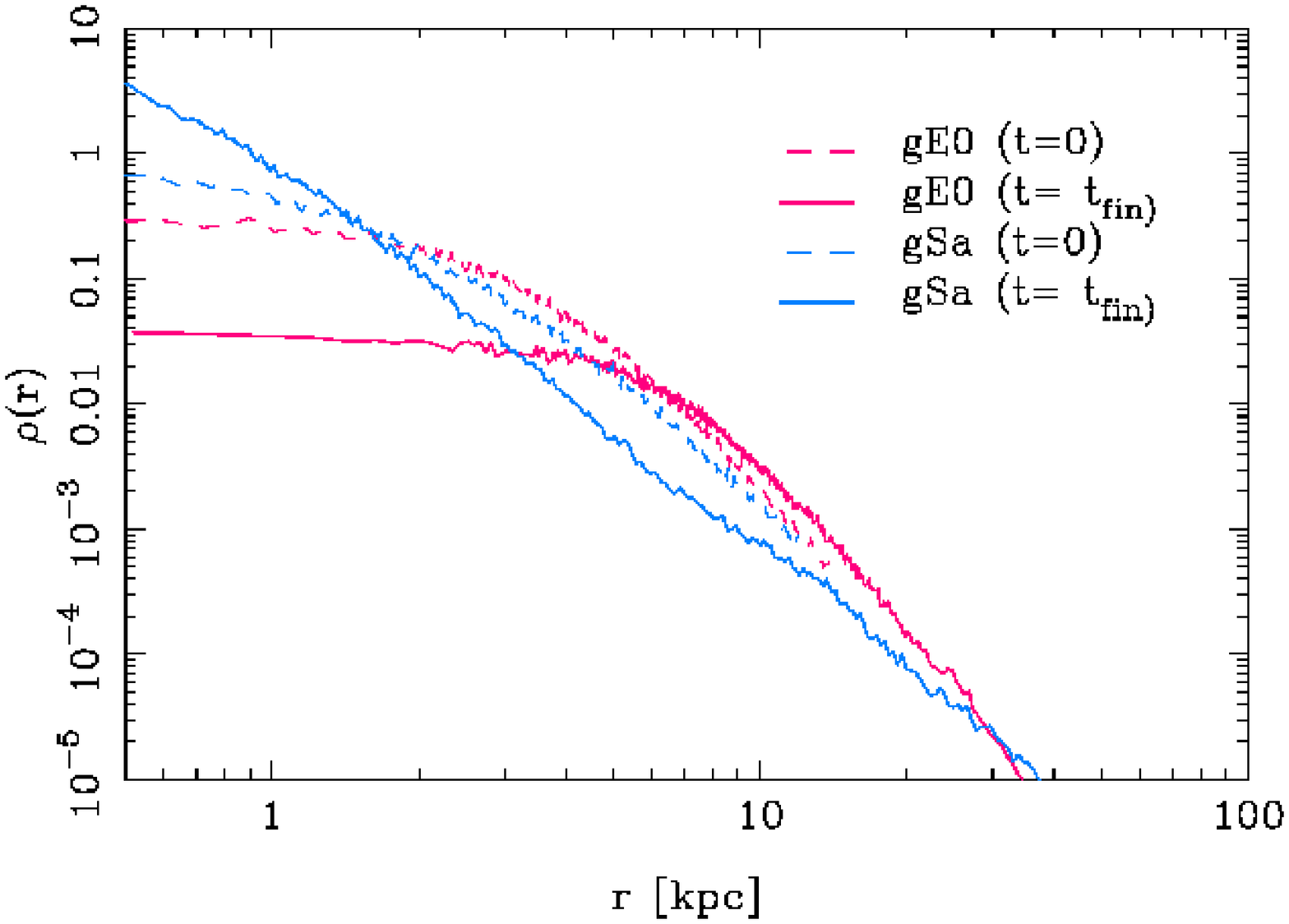}
   \caption{Volume density profile of the stars initially belonging to the elliptical galaxy (red curves) and to the spiral Sa (blue curves). The dashed lines show the initial (i.e. at the beginning of the simulation) density profiles, while the solid lines show the final distributions. This plot refers to the 'fiducial' dissipative merger presented in Sect.\ref{e0sa}. }
              \label{fig5app}%
 \end{minipage}\\
 \ \hspace{1mm} \
 \begin{minipage}[t]{9cm}
   \centering
    \includegraphics[width=5cm,angle=270]{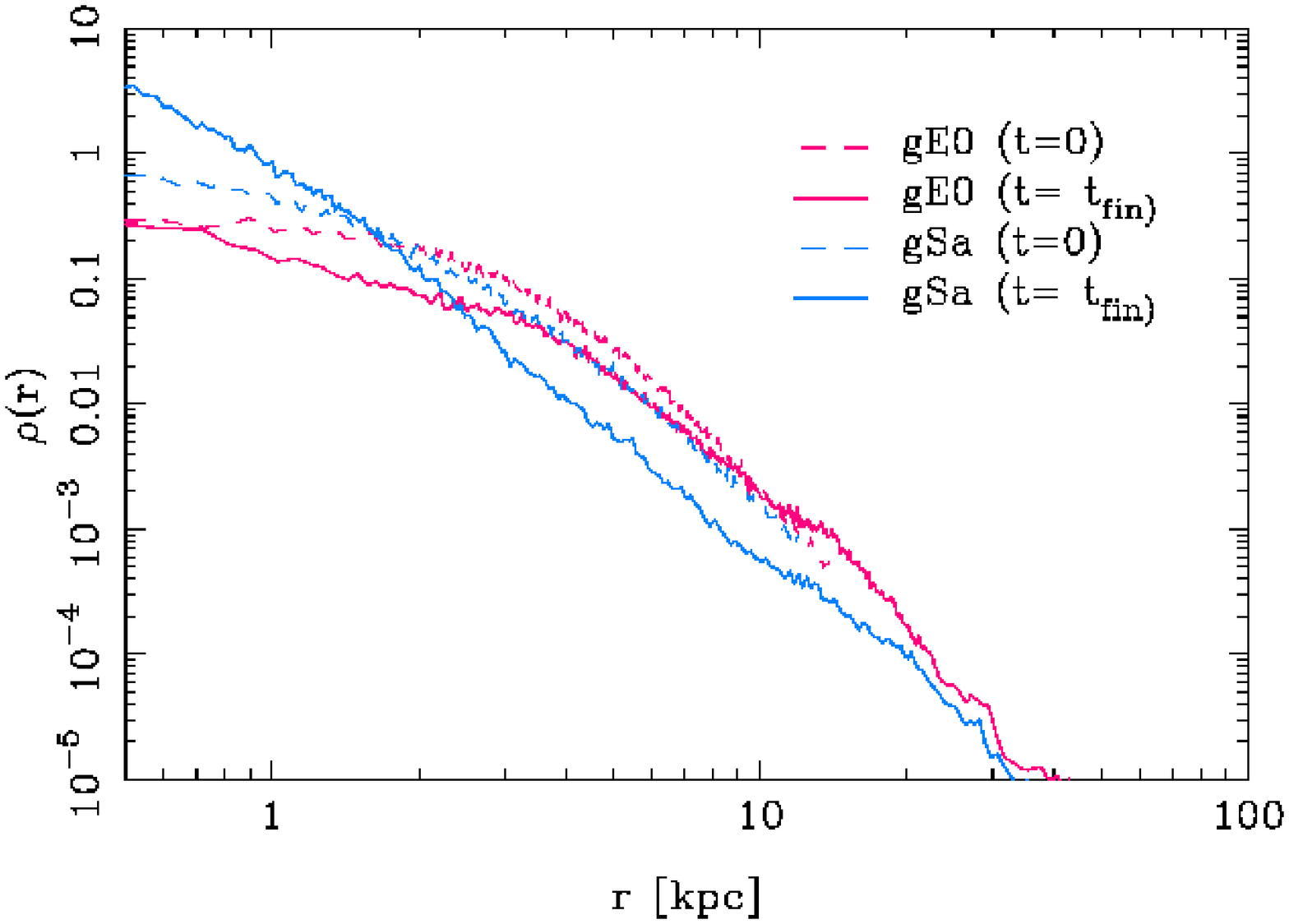}
   \caption{Same as Fig.\ref{fig5app}, but for the coplanar merger, whose velocity field is shown in Fig.\ref{fig1app}.} 
              \label{fig6app}%
 \end{minipage}\\
 \ \hspace{1mm} \
\begin{minipage}[t]{9cm}
   \centering
 \includegraphics[width=5cm,angle=270]{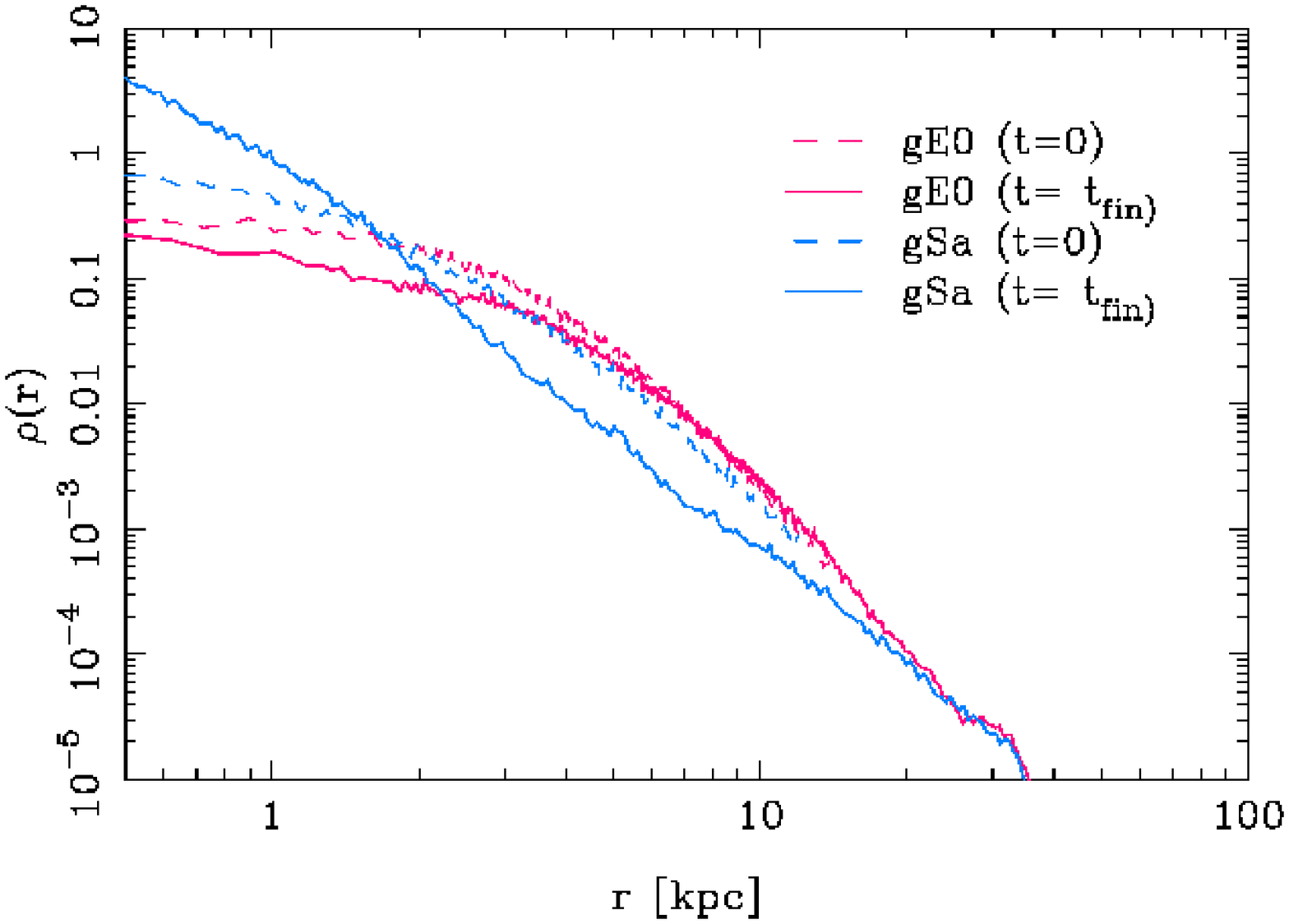}
   \caption{Same as Fig.\ref{fig5app}, but for the coplanar merger, whose velocity field is shown in Fig.\ref{fig2app}.}
              \label{fig7app}
 \end{minipage}
    \end{figure}

In Sect.\ref{old}, we have seen that the emergence of a counter-rotating core in the 'fiducial' merger remnant depends on the fact that the inner region of the former spiral Sa preserves part of its initial spin, while the outer regions, as well as the elliptical galaxy, acquire part of the orbital angular momentum. Evidently, the velocity field of the remnant depends on the final spatial distribution of the two systems (i.e. of stars initially belonging to the elliptical and of the stars initially belonging to the spiral galaxy). As we showed in Fig.\ref{zone}, during the interaction, the elliptical galaxy acquires part of the orbital angular momentum. This results into an overall expansion of the system, as shown in Fig.\ref{fig5app}, where the initial and final volume density distribution of the elliptical system are compared. Even if preserving the initial spin orientation, the inner regions of the spiral lose  part of their angular momentum, acquiring part of the orbital one. This leads to a shrinking of the inner stellar distribution of the spiral system (see Fig.\ref{fig5app}, blue curves).  In particular, it can be seen that stars initially belonging to the spiral dominate the stellar distribution at a distance $r\le 3$ kpc from the center of the remnant. \\
Analysing the other two coplanar mergers discussed in Sect. \ref{app1}, we found that, also in this case, old stars formerly in the spiral Sa galaxy dominate the volume density profile in the inner parts of the remnant galaxies, at distances $r\le 2- 3$ kpc from the center (see Figs.\ref{fig6app} and \ref{fig7app}). But the reader can notice that, while the final density profiles of stars initially belonging to the elliptical and to the spiral galaxy are quite similar in these two cases, the velocity fields shown in Figs.\ref{fig1app} and \ref{fig2app} are, in turn, quite different. In particular, the extension of the counter-rotating region is, in Fig.\ref{fig1app} about twice the one shown in Fig.\ref{fig2app}. In other words,  the two coplanar remnants presented in Sect. \ref{app1} have different velocity fields (in terms of extension and amplitude of the counter-rotating region) even if they have remarkably similar density profiles for the stars in the former elliptical and in the former spiral Sa. 
This means that the relative spatial distribution of stars in the former elliptical and in the former spiral is not the main parameter that determines the principal features of the counter-rotating region, but it is rather the orbital parameters.
\end{appendix}

\end{document}